\documentclass[]{spie}  
\usepackage[]{graphicx}
\usepackage{color}
\usepackage{amsmath}
\definecolor{blue}{rgb}{0.1,0.1,0.6}
\definecolor{orange}{rgb}{0.74,.35,0.099}
\definecolor{pale}{rgb}{0.90,0.90,0.95}
\definecolor{red}{rgb}{1.0,0.0,0.0}

\title{Gemini Planet Imager Observational Calibrations VII: On-Sky Polarimetric Performance of the Gemini Planet Imager} 

\author{Sloane J. Wiktorowicz\supit{a}\footnote{\hspace{5pt}NASA Sagan Fellow}, Max Millar-Blanchaer\supit{b}, Marshall D. Perrin\supit{c}, James R. Graham\supit{d}, Michael P. Fitzgerald\supit{e}, J\'{e}r\^{o}me Maire\supit{f}, Patrick Ingraham\supit{g,h}, Dmitry Savransky\supit{i}, Bruce A. Macintosh\supit{j,k}, Sandrine J. Thomas\supit{l}, Jeffrey K. Chilcote\supit{e}, Zachary H. Draper\supit{m}, Inseok Song\supit{n}, Andrew Cardwell\supit{o}, Stephen J. Goodsell\supit{p}, Markus Hartung\supit{o}, Pascale Hibon\supit{o}, Fredrik Rantakyr\"o\supit{o}, Naru Sadakuni\supit{o}, \& the GPI team
\skiplinehalf
\supit{a}Department of Astronomy, UC Santa Cruz, 1156 High Street, Santa Cruz, CA 95064, USA; \\
\supit{b}Department of Astronomy \& Astrophysics, University of Toronto, Toronto, ON, Canada; \\\
\supit{c}Space Telescope Science Institute, 3700 San Martin Drive, Baltimore, MD 21218, USA; \\\
\supit{d}Department of Astronomy, UC Berkeley, Berkeley, CA 94720, USA; \\\
\supit{e}Department of Physics \& Astronomy, UCLA, Los Angeles, CA 90095, USA; \\\
\supit{f}Dunlap Institute for Astronomy \& Astrophysics, Univ. of Toronto, Toronto, ON, Canada; \\\
\supit{g}Kavli Institute for Particle Astrophysics \& Cosmology, Stanford University, CA, USA; \\\
\supit{h}Departement de Physique, Universit\'e de Montr\'eal, Montr\'eal, PQ, Canada; \\\
\supit{i}Sibley School of Mechanical \& Aerospace Engineering, Cornell University, Ithaca, NY, USA; \\\
\supit{j}Lawrence Livermore National Lab, 7000 East Avenue, Livermore, CA 94551, USA; \\\
\supit{k}Department of Physics \& Astrophysics, Stanford University, Stanford, CA 94305, USA; \\\
\supit{l}UARC/NASA Ames Research Center, Moffett Field, CA 94035, USA; \\\
\supit{m}Department of Physics \& Astronomy, University of Victoria, Victoria, BC, Canada; \\\
\supit{n}Department of Physics \& Astronomy, University of Georgia, Athens, GA 30602, USA; \\\
\supit{o}Gemini Observatory, Southern Operations Center, La Serena, CL; \\\
\supit{p}Gemini Observatory, Northern Operations Center, Hilo, HI, 96720, USA
}

\authorinfo{Further author information: (Send correspondence to S.J.W.)\\S.J.W.: E-mail: sloanew@ucolick.org, Telephone: 1 831 502 7258}

\begin{document} 
  \maketitle 

\begin{abstract}
We present on-sky polarimetric observations with the Gemini Planet Imager (GPI) obtained at straight Cassegrain focus on the Gemini South 8-m telescope. Observations of polarimetric calibrator stars, ranging from nearly unpolarized to strongly polarized, enable determination of the combined telescope and instrumental polarization. We find the conversion of Stokes $I$ to linear and circular instrumental polarization in the instrument frame to be $I \rightarrow (Q_{\rm IP}, U_{\rm IP}, P_{\rm IP}, V_{\rm IP}) = (-0.037 \pm 0.010\%, +0.4338 \pm 0.0075\%, 0.4354 \pm 0.0075\%, -6.64 \pm 0.56\%)$. Such precise measurement of instrumental polarization enables $\sim 0.1\%$ absolute accuracy in measurements of linear polarization, which together with GPI's high contrast will allow GPI to explore scattered light from circumstellar disk in unprecedented detail, conduct observations of a range of other astronomical bodies, and potentially even study polarized thermal emission from young exoplanets. Observations of unpolarized standard stars also let us quantify how well GPI's differential polarimetry mode can suppress the stellar PSF halo. We show that GPI polarimetry achieves cancellation of unpolarized starlight by factors of 100-200, reaching the photon noise limit for sensitivity to circumstellar scattered light for all but the smallest separations at which the calibration for instrumental polarization currently sets the limit. 
\end{abstract}



\section{INTRODUCTION}
\label{sec:intro}  

The Gemini Planet Imager (GPI) obtained first light at the Gemini South 8-m telescope in November 2013\cite{Macintosh14}, and its first on-sky polarimetric observations occurred in December 2013. The polarimetric mode of this instrument is primarily intended to spatially resolve scattered light from circumstellar dust, in particular from protoplanetary and debris disks\cite{Perrin10}. Additionally, GPI will enable studies ranging from Titan's atmospheric properties to outflows from high mass stars. For a description of the design and implementation of GPI's polarimetry mode, data reduction methods, and the first polarimetric science results from GPI, see M. Perrin et al. (2014, in preparation). An integral part of data reduction involves correction for instrumental polarization.

Instrumental polarization bias arises from crosstalk between Stokes parameters ${I,Q,U,V}$ as light is transferred through the combined telescope and instrument system, which can be represented by the off-diagonal elements in the Mueller matrix describing the system's polarization response. The polarization at straight Cassegrain focus imparted by the telescope itself is expected to be lower than that introduced by GPI. Indeed, the dominant contributor to system polarization is expected to be GPI's own optics, which include 12 reflections at varying angles of incidence prior to GPI's halfwave plate. Based on numerical ray tracing, GPI is likely to have of order $1\%$ instrumental polarization. In contrast, telescope polarization at straight Cassegrain focus is typically measured to be of order $0.01\%$\cite{Wiktorowicz08, Wiktorowicz09}.

GPI is required to be operable on all ports of the Gemini Instrument Support Structure (ISS), which include the straight-through bottom port as well as the side ports which receive light via a $45^{\circ}$ fold mirror, M3. For the side ports, this $45^{\circ}$ fold is expected to introduce enhanced telescope polarization, which is expected to be of order 1\% based on modeling of Gemini's multilayer protected silver coating \cite{Perrin10}. However, it is expected that GPI observing runs will make preferential use of the bottom (straight-through) port, which improves both telescope polarization and overall contrast by minimizing reflections. Therefore, we do not consider the contribution of the M3 fold to telescope polarization. In this paper, we describe on-sky observations of polarization calibrator stars to assess the performance of GPI's polarimetry mode. We discuss measurements of instrumental polarization in \S2 and \S3, and \S4 illustrates GPI's capabilities for suppression of unpolarized stellar PSF halos.

\section{METHODS FOR MEASURING INSTRUMENTAL POLARIZATION}

To derive measurements for GPI's instrumental polarization, we first set up a calculation framework of Mueller matrices describing GPI's sensitivity to incident polarized starlight, and we then apply that framework to observations of polarized and unpolarized standard stars observed during the commissioning of GPI's polarimetry mode in December 2013.

\subsection{Mueller Matrices Representing GPI Polarimetry Observations}
\label{sec:mueller}

GPI always operates in angular differential imaging (ADI) mode, where the alt-az telescope's Cassegrain rotator is fixed in angle so the pupil is static in the instrument frame. This increases the accuracy of quasi-static speckle subtraction when searching for circumstellar disks or planets. Therefore, both the pupil and the telescope/instrument polarization (IP) vector rotate with respect to the sky according to the observed parallactic angle $\phi_{\rm par}$. However, stellar polarization, referenced to celestial north, does not rotate with parallactic angle, which allows stellar polarization and IP to be separated.

Using the Mueller matrix formalism, the Stokes vector polarization state $S_{\rm HWP}$ of light reaching GPI's half-wave plate is the following, assuming an incident Stokes vector $S_{\rm source}$ with total intensity normalized to 1:

	\begin{eqnarray}
S_{\rm HWP} & = & M_{\rm IP} T_{\phi} S_{\rm source} =
\left(
\begin{array}{cccc}
  1 			& 0 	& 0 	& 0 \\
  Q_{\rm IP} 	& 1	& 0	& 0 \\
  U_{\rm IP} 	& 0	& 1	& 0 \\
  V_{\rm IP}	& 0	& 0	& 1
\end{array}
\right) \label{eq:1}
\left(
\begin{array}{cccc}
  1 & 0 			& 0 			& 0 \\
  0 & \cos{2\phi}		& \sin{2\phi}	& 0 \\
  0 & -\sin{2\phi}	& \cos{2\phi}	& 0 \\
  0 & 0			& 0			& 1
\end{array}
\right)
\left(
\begin{array}{c}
  1 \\
  Q \\
  U \\
  0
\end{array}
\right) \\
& = &
\left(
\begin{array}{c}
  1 \\
  Q_{\rm IP} + Q \cos{2\phi} + U \sin{2\phi} \\
  U_{\rm IP} - Q \sin{2\phi} + U \cos{2\phi} \\
  V_{\rm IP}
\end{array}
\right) \nonumber.
	\end{eqnarray}
	
\noindent Here, $T_\phi$ is the rotation matrix to rotate stellar polarization into the instrumental reference frame (where $\phi$ is related to parallactic angle, see below), and the normalized stellar Stokes parameters are taken to be $Q/I$, $U/I$. We assume stellar $V/I=0$. For the sake of brevity, we drop the denominator $I$ for the rest of this manuscript. The subscript ``IP" indicates the combination of telescope and instrumental polarization, which cannot be distinguished from each other due to their identical modulation with parallactic angle. Optical\label{sec:zemax}, Zemax modeling of the reflections and nominal phase retardances from mirror coatings in GPI suggest $P_{\rm IP} = \sqrt{Q_{\rm IP}^2 + U_{\rm IP}^2} \sim 1\%$ and $V_{\rm IP} \sim 10\%$. Each of these quantities is much larger than the expected telescope polarization, for which measurements are typically of order $0.01\%$\cite{Wiktorowicz08, Wiktorowicz09}.

To minimize the number of free parameters in fitting on-sky data, we only model the first column of the telescope/instrumental polarization Mueller matrix. The $I \rightarrow (Q, U, V)$ matrix elements are therefore indicated by $(Q_{\rm IP}, U_{\rm IP}, V_{\rm IP})$, respectively. In the instrumental reference frame (Equation \ref{eq:1}), stellar polarization $(Q, U)$ rotates with parallactic angle while instrumental polarization $(Q_{\rm IP}, U_{\rm IP}$) is constant. After the stellar beam enters GPI, it propagates through the various subsystems and experiences multiple reflections and phase retardance. By the time the beam reaches the halfwave plate, it has experienced significant instrumental polarization. The Stokes vector state of the beam reaching the integral field spectrograph (IFS) detector is given by the following:

	\begin{eqnarray}
S_{\rm IFS} & = & M_{\rm Wol} T_{-\psi} M_{\rm HWP}(\lambda) T_{\psi} S_{\rm HWP} =
\frac{1}{2} \left(
\begin{array}{cccc}
  1 		& \pm 1 	& 0 	& 0 \\
  \pm 1	& 1		& 0	& 0 \\
  0 		& 0		& 0	& 0 \\
  0 		& 0		& 0	& 0
\end{array}
\right)
\left(
\begin{array}{cccc}
  1 & 0 			& 0 			& 0 \\
  0 & \cos{2\psi}		& -\sin{2\psi}	& 0 \\
  0 & \sin{2\psi}		& \cos{2\psi}	& 0 \\
  0 & 0			& 0			& 1
\end{array}
\right) \label{eq:2} \\ \nonumber
& \times &
\left(
\begin{array}{cccc}
  1 	& 0 	& 0 				& 0 \\
  0 	& 1	& 0				& 0 \\
  0	& 0	& m_{3,3}(\lambda)	& m_{3,4}(\lambda) \\
  0	& 0	& m_{4,3}(\lambda)	& m_{4,4}(\lambda)
\end{array}
\right)
\left(
\begin{array}{cccc}
  1 & 0 			& 0 			& 0 \\
  0 & \cos{2\psi}		& \sin{2\psi}	& 0 \\
  0 & -\sin{2\psi}	& \cos{2\psi}	& 0 \\
  0 & 0			& 0			& 1
\end{array}
\right)
\left(
\begin{array}{c}
  1 \\
  Q_{\rm IP} + Q \cos{2\phi} + U \sin{2\phi} \\
  U_{\rm IP} - Q \sin{2\phi} + U \cos{2\phi} \\
  V_{\rm IP}
\end{array}
\right)
	\end{eqnarray}

\noindent Here, $M_{\rm Wol}$ is the Mueller matrix for the MgF$_2$/SiO$_2$ Wollaston prism, which bifurcates the beam into opposite polarization states denoted by ``$\pm$". Given the known orientation of the Wollaston prism within the GPI IFS, the true orientation of the Wollaston axis with respect to the plane of the sky is given by $\phi = \phi_{\rm par} + 90^\circ - 18.5^\circ$, where parallactic angle is given by $\phi_{\rm par}$. The architecture of the integral field polarimetry system motivates the 18.5$^\circ$ orientation for the Wollaston prism (M. Perrin et al. 2014, in preparation), and the 90$^\circ$ offset simply reflects the arbitrary convention defining the reference angle for astronomical Stokes $+Q$. Laboratory tests, using a linear polarizer and quarterwave plate to inject beams of known Stokes state into GPI, indicate that the halfwave plate fast axis is oriented $29.14^\circ$ with respect to the rotation stage\cite{Wiktorowicz12}. The orientation of the halfwave plate's optical axis projected on the plane of the sky is therefore given by $\psi = -(\psi_{\rm cmd} + 29.14^\circ)$, where $\psi_{\rm cmd}$ is the commanded orientation and is a multiple of $22.5^\circ$ during a typical observing sequence.

The Mueller matrix of the GPI halfwave plate $M_{\rm HWP}(\lambda)$ is approximately but not perfectly achromatic. Table \ref{tab:hwp} lists Mueller matrix parameters for the halfwave plate in $Y$, $J$, $H$, $K1$, and $K2$ bands, which are updated from a previous report of the $H$ band retardance\cite{Wiktorowicz12}. The updated retardance values were found by fitting vendor-provided retardances (measured at 0.894, 1.194, 1.242, 1.529 and 1.555~$\mu$m on a witness sample fabricated alongside our halfwave plate) to the retardances calculated using the known birefringence values of MgF$_2$ and crystalline SiO$_2$. The thicknesses of the two layers were left as free parameters and adjusted to fit the measurements. The best fit thicknesses derived via this method were within 0.25\% of the nominal design thicknesses from the vendor. These best fit thicknesses were then recombined with the birefringence to determine the retardance of the waveplate for the complete GPI wavelength range from $0.9$ to $2.5~\mu$m.  The known transmission of each of GPI's filters was then used to perform a weighted mean of the retardance across each band. 

\clearpage 

For each halfwave plate orientation $\psi$, the Wollaston prism and lenslet array generate a pair of spots on the IFS that encode the fractional polarization, which is effectively the measurement of a single linear Stokes parameter:

	\begin{eqnarray}
P_{\rm frac}(\psi) = \frac{I^+ - I^-}{I^+ + I^-}.
	\label{eq:4}
	\end{eqnarray}

\noindent Here, $I^\pm$ are given by the first element in the $4\times1$ matrix $S_{\rm IFS}$ in Equation \ref{eq:2} for the two opposite polarization outputs of the Wollaston, respectively. Taking into account parallactic angle and GPI's halfwave plate orientation, $P_{\rm frac}(\psi)$ is given by the following:

	\begin{eqnarray}
P_{\rm frac}(\psi) & = & (\tfrac{1-m_{3,3}}{2}) [\sin{4\psi} (U_{\rm IP} - Q\sin{2\phi} + U \cos{2\phi}) + (\cos{4\psi} + \tfrac{1+m_{3,3}}{1-m_{3,3}}) (Q_{\rm IP} + Q \cos{2\phi} + U \sin{2\phi}) ] \label{eq:3} \\
& - & m_{3, 4} V_{\rm IP} \sin{2\psi} . \nonumber
	\end{eqnarray}
	
\noindent It can be seen from Equation \ref{eq:3} that, as measured in GPI's instrumental reference frame, only the astrophysical polarization components $Q$ and $U$ are modulated by parallactic angle; the instrumental polarization is independent of parallactic rotation.  Additionally, while $Q$, $U$, $Q_{\rm IP}$, and $U_{\rm IP}$ are modulated by the halfwave plate at $4\psi$, $V_{\rm IP}$ is modulated at $2\psi$. Thus, the circular IP will manifest as a slow modulation of the halfwave plate and will be constant with parallactic angle. Based on Zemax modeling and laboratory measurements\cite{Wiktorowicz12}, GPI is expected to exhibit instrumental circular polarization of order $1\%$ in on-sky observations.

\begin{table}[t]
\caption{GPI halfwave plate retardances.} 
\label{tab:hwp}
\begin{center}       
\begin{tabular}{|c|c|c|c|c|} 
\hline
\rule[-1ex]{0pt}{3.5ex}  Band	& Central Wavelength ($\mu m$) & Retardance (waves)	& $m_{3,3} = m_{4,4}$	& $m_{3,4} = -m_{4,3}$  \\
\hline
\rule[-1ex]{0pt}{3.5ex}  $Y$	& 1.045 & 0.4959				& $-0.9997$			& $+0.0258$		 \\
\rule[-1ex]{0pt}{3.5ex}  $J$	& 1.232 & 0.4863				& $-0.9963$			& $+0.0860$		\\
\rule[-1ex]{0pt}{3.5ex}  $H$	& 1.647 & 0.4819				& $-0.9935$			& $+0.1135$		 \\
\rule[-1ex]{0pt}{3.5ex}  $K1$	& 2.045 & 0.4945				& $-0.9994$			& $+0.0346$		 \\
\rule[-1ex]{0pt}{3.5ex}  $K2$	& 2.255 & 0.5085				& $-0.9986$			& $-0.0534$		 \\
\hline 
\end{tabular}
\end{center}
\end{table}

\subsection{Measurements and Data Reduction}
\label{sec:datred}

From combined $V$ band polarimetric catalogs\cite{Heiles00}, we determine suitable nearly unpolarized and strongly polarized calibrator stars to observe with GPI. The first stellar polarimetric observations taken with GPI occurred on the night of December 13, 2013 UT and consisted of the nearly unpolarized star HD 12759, the strongly polarized stars HD 77581 and HD 78344, and the circumstellar disk hosts $\beta$ Pic, HD 15115, and HR 4796A. See Table \ref{tab:stars} for literature measurements for the polarization of these stars, where available, with uncertainties in the last two digits given in parentheses.  Clearly, the wavelength dependence of interstellar polarization (\S \ref{sec:int}) suggests that our near-infrared measurements will differ from those in the optical.

\begin{table}[t]
\caption{Observed Stars. $P$ and $\Theta$ give the observed stellar percent polarization and position angle in the wavelength band indicated and from the references cited in the right column. To our knowledge, no polarimetric observations of the host stars HD 15115 and HR 4796A themselves have yet been published.} 
\label{tab:stars}
\begin{center}       
\begin{tabular}{|c|c|c|c|c|c|c|c|} 
\hline
\rule[-1ex]{0pt}{3.5ex}  Name		& Alt. Name	& R.A. (J2000)	& Dec. (J2000)		& $P (\%)$	& $\Theta (^\circ)$	& Band	& Ref.  \\
\hline
\rule[-1ex]{0pt}{3.5ex}  HD 12759	& GJ 9069AB	& 02 03 55.25	& $-$45 24 46.5	& $< 0.021$	& $-$			& $V$		& \cite{Schroeder76, Heiles00} \\
\rule[-1ex]{0pt}{3.5ex}  HD 15115	& HIP 11360	& 02 26 16.24	& $+$06 17 33.2	& $-$		& $-$			& $-$		& $-$ \\
\rule[-1ex]{0pt}{3.5ex}  $\beta$ Pic	& HD 39060	& 05 47 17.09	& $-$51 03 59.4	& $0.089(24)$	& $65.2(8.0)$		& 700-900 nm	& \cite{Tinbergen82} \\
\rule[-1ex]{0pt}{3.5ex}  HD 77581	& Vela X-1		& 09 02 06.86	& $-$40 33 16.9	& $3.739(62)$	& $82.9(0.5)$		& $V$		& \cite{Hall58, Mathewson70, Klare77, Heiles00} \\
\rule[-1ex]{0pt}{3.5ex}  HD 78344	& HIP 44647	& 09 05 51.33	& $-$47 46 06.8	& $5.56(10)$	& $169.7(0.5)$		& $V$		& \cite{Klare77, Heiles00} \\
\rule[-1ex]{0pt}{3.5ex}  HR 4796A	& $-$		& 12 36 01.03	& $-$39 52 10.2	& $-$		& $-$			& $-$		& $-$ \\
\hline 
\end{tabular}
\end{center}
\end{table}

Observations of the strongly polarized stars were performed in unocculted/open loop mode, where the light from the star was spread over many IFS lenslets. The nearly unpolarized binary star HD 12759 (= GJ 9069AB, separation measured to be $1.450 \pm 0.051"$) was observed in $H$ band in both occulted/closed loop and unocculted/open loop modes, and the disk host stars were observed in $H$ band in occulted/closed loop mode (with HR 4796A also observed in $K1$ band). Unocculted/open loop observations of HD 12759, HD 77581, and HD 78344 were taken in $J$, $H$, and $K1$ bands to assess instrumental performance with wavelength of observation. The raw data were reduced to the point of producing oppositely polarized datacubes using the GPI Data Reduction Pipeline \cite{Maire10,perrinthis} but not yet combined as a sequence.  The quantity $P_{\rm frac}(\psi)$ from Equation \ref{eq:4} was calculated for each frame. This calculation, normalized by the total flux, is a necessity for the open loop observations. This is because the morphology and position of the stellar PSF varies from frame to frame, and more importantly, across different orientations of the halfwave plate.

\begin{figure}
   \begin{center}
   \begin{tabular}{c}
   \includegraphics[height=8.8cm]{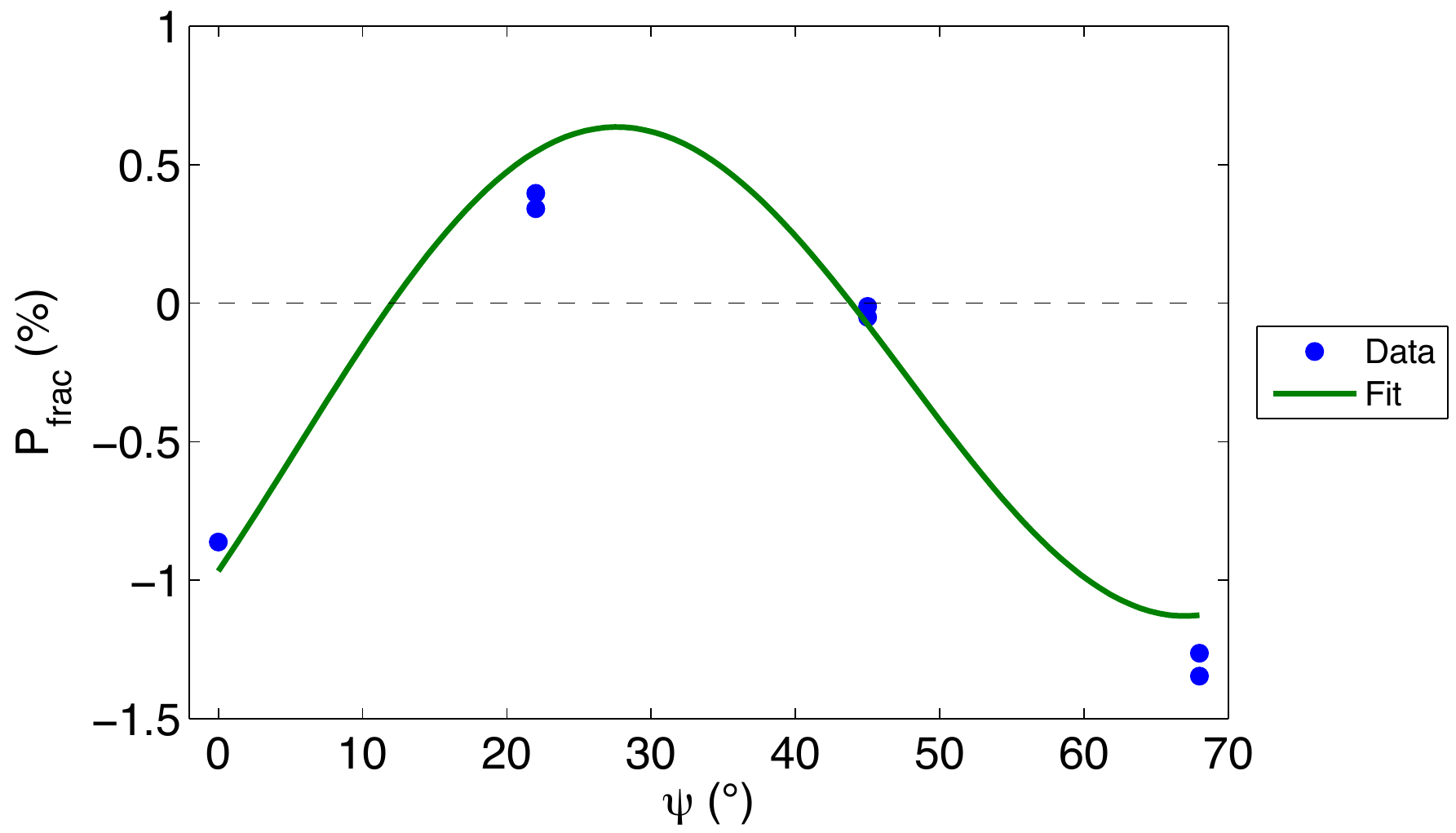}
   \end{tabular}
   \end{center}
   \caption[example] 
   { \label{fig:pvshwp} 
$H$ band, closed loop, occulted observations of the polarized standard HD 77581. Fractional polarization $P_{\rm frac}$ is shown as a function of halfwave plate orientation $\psi$. Data are given by points, and the best fit to Equation \ref{eq:3} is given by the curve. The nonzero, instrumental circular polarization of GPI ($V_{\rm IP}$) is evidenced by the nonzero mean value of the data. That is, rotation of the halfwave plate by $\Delta \psi = 45^\circ$ changes the sign of measured $P_{\rm frac}$, but preserves its magnitude, only for $V_{\rm IP} = 0$.}
   \end{figure}

Aperture polarimetry was performed for each frame at the center of the PSF. For observations in occulted mode, we assume that the polarization of the flux just outside of the occulter is representative of the true stellar polarization. This assumption will be directly tested by both occulted and unocculted observations of HD 12759 (see \S \ref{sec:occ}). A nonlinear least-squares procedure (specifically Gauss-Newton minimization) was employed to determine $Q$, $U$, $Q_{\rm IP}$, $U_{\rm IP}$, and $V_{\rm IP}$ from Equation \ref{eq:3}. It is not required that the measurements of IP be obtained on truly unpolarized stars, as IP modulates with parallactic angle while stellar polarization does not. With a sufficient range in observed parallactic angle, any non-variable star may theoretically be suitable for measurement of IP. The result of this fitting procedure is shown in Figure \ref{fig:pvshwp}, which illustrates $H$ band observations of HD 77581 as a function of $\psi$.

\section{RESULTS}
\subsection{Telescope/Instrumental Polarization}

\begin{table}[t]
\caption{Observed Stellar Polarization.  $+Q$ is referenced to celestial north and $\Theta$ is defined to be positive east of north. See text for the meaning of the AO column.} 
\label{tab:stellar_pol}
\begin{center}       
\begin{tabular}{|c|c|c|c|c|c|c|c|c|} 
\hline
\rule[-1ex]{0pt}{3.5ex}  Star			& Band	& AO			& Apodizer	& $\Delta \phi_{\rm par}$	& $Q (\%)$		& $U (\%)$		& $P (\%)$	& $\Theta (^\circ)$  \\
\hline
\hline
\rule[-1ex]{0pt}{3.5ex}  GJ 9069A		& $H$	& CO		& In			& $15.8^\circ$			& $-0.016(21)$		& $+0.036(23)$	& 0.039(24)	& 57(18) \\
\rule[-1ex]{0pt}{3.5ex}  $\cdots$		& $J$	& OU	& Out		& $3.2^\circ$			& $-0.073(44)$		& $-0.076(44)$		& 0.106(46)	& 113(13) \\
\rule[-1ex]{0pt}{3.5ex}  $\cdots$		& $H$	& $\cdots$	& $\cdots$	& $4.9^\circ$			& $-0.32(29)$		& $-0.34(29)$		& 0.47(32)	& 113(21) \\
\rule[-1ex]{0pt}{3.5ex}  $\cdots$		& $K1$	& $\cdots$	& $\cdots$	& $3.0^\circ$			& $-0.27(20)$		& $+0.06(20)$		& 0.28(22)	& 83(25) \\

\hline
\rule[-1ex]{0pt}{3.5ex}  GJ 9069B		& $H$	& CO		& In			& $15.8^\circ$			& $+0.29(26)$		& $+0.02(28)$		& 0.29(33)	& 2(37) \\

\hline
\rule[-1ex]{0pt}{3.5ex}  HD 77581		& $H$	& CO		& In			& $2.1^\circ$			& $-1.066(44)$		& $+0.153(57)$	& 1.077(45)	& 85.9(1.5) \\
\rule[-1ex]{0pt}{3.5ex}  $\cdots$		& $J$	& OU		& In			& $1.2^\circ$			& $-1.889(62)$		& $+0.128(63)$	& 1.893(62)	& 88.06(95) \\
\rule[-1ex]{0pt}{3.5ex}  $\cdots$		& $H$	& $\cdots$	& $\cdots$	& $1.3^\circ$			& $-0.983(34)$		& $+0.237(35)$	& 1.012(35)	& 83.21(99) \\
\rule[-1ex]{0pt}{3.5ex}  $\cdots$		& $K1$	& $\cdots$	& $\cdots$	& $1.2^\circ$			& $-0.359(72)$		& $-0.058(73)$		& 0.363(73)	& 94.6(5.9) \\

\hline
\rule[-1ex]{0pt}{3.5ex}  HD 78344		& $J$	& OU		& In			& $1.8^\circ$			& $+2.42(33)$		& $-1.18(33)$		& 2.70(33)	& 167.0(3.5) \\

\rule[-1ex]{0pt}{3.5ex}  $\cdots$		& $H$	& $\cdots$	& $\cdots$	& $1.8^\circ$			& $+1.94(19)$		& $+0.15(19)$		& 1.94(19)	& 2.2(2.8) \\
\rule[-1ex]{0pt}{3.5ex}  $\cdots$		& $K1$	& $\cdots$	& $\cdots$	& $1.9^\circ$			& $+1.17(17)$		& $-0.25(17)$		& 1.19(17)	& 173.9(4.2) \\

\hline
\rule[-1ex]{0pt}{3.5ex}  $\beta$ Pic		& $H$	& CO		& In			& $91.5^\circ$			& $-0.0367(85)$		& $-0.0164(91)$	& 0.0402(87)	& 102.1(6.6) \\

\hline
\rule[-1ex]{0pt}{3.5ex}  HD 15115		& $H$	& CO		& In			& $12.5^\circ$			& $-0.003(11)$		& $-0.007(12)$	& 0.008(17)	& 123(78) \\

\hline
\rule[-1ex]{0pt}{3.5ex}  HR 4796A		& $H$	& CO		& In			& $2.1^\circ$			& $+0.122(24)$		& $+0.132(28)$	& 0.180(27)	& 23.6(4.2) \\
\rule[-1ex]{0pt}{3.5ex}  $\cdots$		& $K1$	& $\cdots$	& $\cdots$	& $3.1^\circ$			& $+0.231(83)$	& $-0.212(92)$	& 0.313(89)	& 158.7(8.4) \\
\hline 
\end{tabular}
\end{center}
\end{table}

\begin{table}[t]
\caption{Observed Instrumental Polarization. $\pm Q_{\rm IP}$ are measured with respect to the Wollaston prism axes in GPI.} 
\label{tab:ip}
\begin{center}       
\begin{tabular}{|c|c|c|c|c|c|c|c|c|} 
\hline
\rule[-1ex]{0pt}{3.5ex}  Star			& Band	& AO			& Apodizer	& $\Delta \phi_{\rm par}$	& $Q_{\rm IP} (\%)$		& $U_{\rm IP} (\%)$		& $P_{\rm IP} (\%)$	& $\Theta_{\rm IP} (^\circ)$  \\
\hline
\rule[-1ex]{0pt}{3.5ex}  IP ($\beta$ Pic)	& $H$	& CO		& In			& $91.5^\circ$			& $-0.037(10)$		& $+0.4338(75)$	& 0.4354(75)	& 47.42(67) \\
\rule[-1ex]{0pt}{3.5ex}  IP (GJ 9069A)	& $\cdots$& $\cdots$	& $\cdots$		& $15.8^\circ$			& $+0.03(74)$		& $+0.62(68)$		& 0.62(93)	& 44(51) \\
\rule[-1ex]{0pt}{3.5ex}  IP (HD 15115)	& $\cdots$& $\cdots$	& $\cdots$		& $12.5^\circ$			& $-0.13(55)$		& $+0.39(57)$		& 0.41(76)	& 54(65) \\
\hline
\end{tabular}
\end{center}
\end{table}

Table \ref{tab:stellar_pol} lists the measured polarizations for the stars in our sample, where the quantities in parentheses indicate the uncertainties in the last two digits. The third column (``AO") lists the adaptive optics mode for each observations, which are either closed loop and occulted (``CO") or open loop and unocculted (``OU").\footnote{ No closed loop, {\it un}occulted observations were taken during the December 2013 observing run, as it is difficult to identify stars of the right apparent magnitude and color to provide enough $I$ band photons for the adaptive optics loop to close, but not so many near-infrared photons as to saturate the IFS.} 

Table~\ref{tab:ip} shows the instrumental polarization, IP, measured from $H$ band observations of $\beta$ Pic.  From Equation \ref{eq:3}, IP is measured to be $I \rightarrow (Q_{\rm IP}, U_{\rm IP}, P_{\rm IP}, V_{\rm IP}) = (-0.037 \pm 0.010\%, +0.4338 \pm 0.0075\%, 0.4354 \pm 0.0075\%, -6.64 \pm 0.56\%)$, which is reasonable given Zemax modeling of the system (\S \ref{sec:zemax}). Measurement of IP from observations of GJ 9069A and HD 15115, taken in identical observing modes as for $\beta$ Pic, are not nearly as precise but are broadly consistent. Ideally, IP would be measured in each bandpass, because it is likely to be a chromatic effect. However, we do not yet have in hand sufficient data on standard stars with enough field rotation to accurately measure IP for bandpasses other than $H$. We therefore account just for the $H$ band instrumental polarization when determining stellar polarization for $JHK1$ observations in Table \ref{tab:stellar_pol}. Measured instrumental polarization of $P_{\rm IP} \sim 0.4\%$ is an order of magnitude larger than measured telescope polarization at the Palomar 5-m\cite{Wiktorowicz08, Wiktorowicz09}; therefore, it is likely that telescope polarization is negligible compared to instrumental polarization, but as a practical matter there is no need to distinguish between the two.

\subsection{Polarimetry of Coronagraphically Occulted Stars}
\label{sec:occ}
Closed loop, occulted observations of the nearly unpolarized star HD 12759 (= GJ 9069AB) were taken in $H$ band with the occulter centered on GJ 9069A (Table \ref{tab:stellar_pol}). Thus, assuming that the polarization of both binary components is identical, the polarization of the occulted GJ 9069A PSF may be compared with that for the unocculted GJ 9069B to gauge the polarimetric effect of the coronagraphic occultation, if any. We find $\Delta (Q, U, P, \Theta) = (-0.31 \pm 0.26\%, 0.02 \pm 0.36\%, 0.31 \pm 0.40, 89^\circ \pm 44^\circ)$, where $\Delta (Q, U) = (Q, U)_{\rm occ} - (Q, U)_{\rm unocc}$ and $\Delta (P, \Theta)$ are calculated from $\Delta (Q, U)$. In addition, both closed loop, occulted and open loop, unocculted observations of the strongly polarized star HD 77581 were taken in $H$ band, and we find $\Delta (Q, U, P, \Theta) = (-0.083 \pm 0.056\%, -0.084 \pm 0.067\%, 0.118 \pm 0.066, 113^\circ \pm 17^\circ)$. The values of $\Delta (Q, U)$ are inconsistent with a null result at only the $0.9\sigma$ level. Therefore, the $95\%$ confidence upper limit to spurious polarization introduced by the occulter is $\Delta (Q, U) < 0.17\%$. It must be stressed that this result is only valid for stars lacking significant scattered light from circumstellar material just outside the occulter spot. This is because the apparent contrast ratio between such close-in material and the star is increased by orders of magnitude due to coronagraphic starlight suppression (see \S \ref{sec:sup}). Therefore, the increased visibility of scattered light from a disk will bias the measurement of stellar polarization. 

The star $\beta$ Pictoris is of course itself a well known circumstellar disk host star. However, in practice the optical depth of the disk-scattered light around $\beta$ Pic is low, and even in coronagraphically occulted images its contribution to the integrated total starlight just outside the occulter is less than that of the PSF halo. Therefore, $\beta$ Pic is an acceptable target for measuring IP, and it is the star from the December 2013 dataset which has the greatest parallactic rotation and therefore the largest modulation lever arm for measuring IP.  Observations of these same data analyzed in another manner to detect the disk will be presented in a later work (M. Millar-Blanchaer et al., in preparation). 

\subsection{Interstellar Polarization}
\label{sec:int}
   \begin{figure}
   \begin{center}
   \begin{tabular}{c}
   \includegraphics[height=8.8cm]{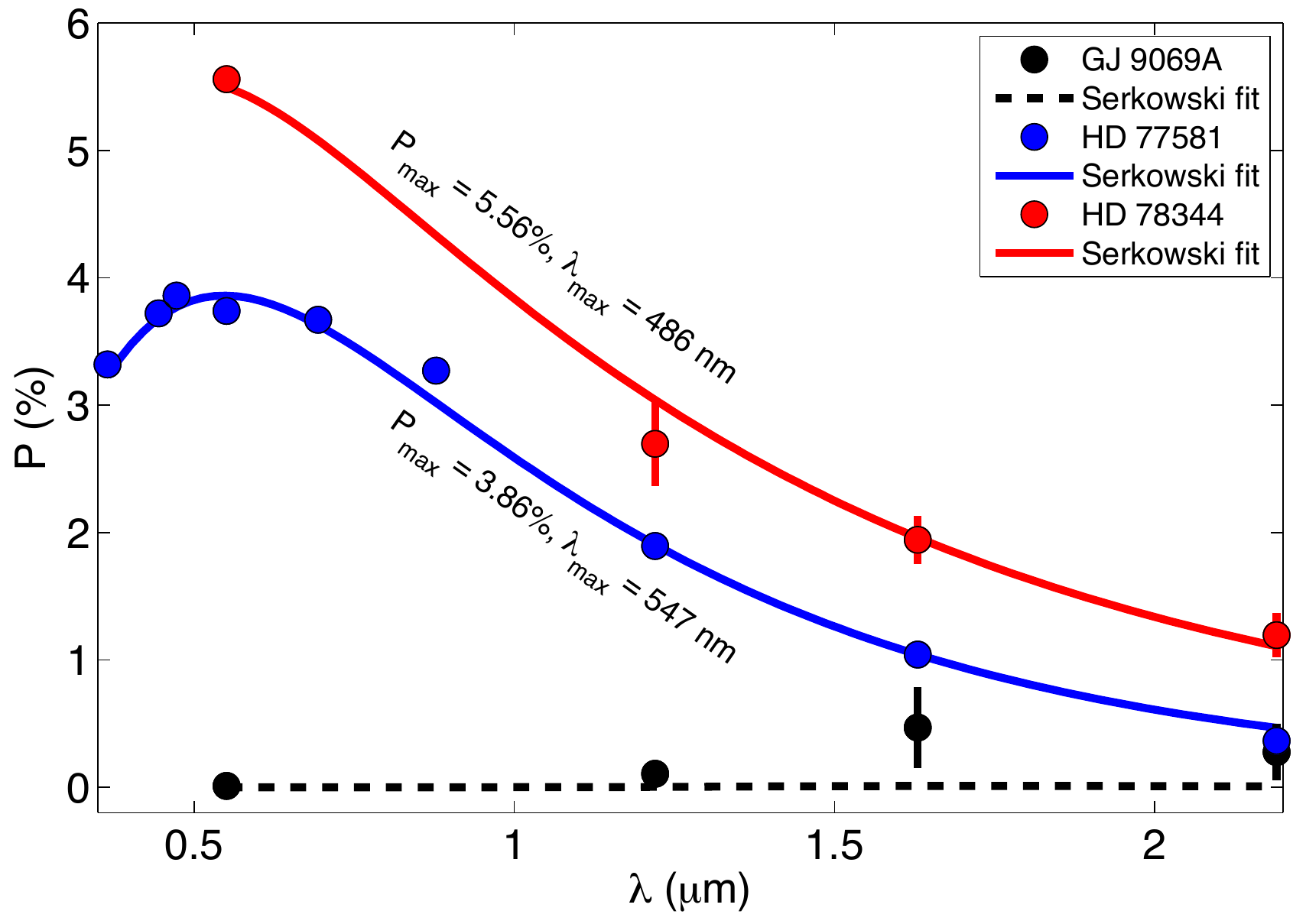}
   \end{tabular}
   \end{center}
   \caption[example] 
   { \label{fig:serk} 
Optical (from the literature, see Table \ref{tab:stars} for references) and near-infrared polarization ($JHK1$ bands from this work), of two strongly polarized stars and one nearly unpolarized star, as a function of wavelength of observation. Multi-wavelength GPI measurements are consistent with interstellar polarization for the strongly polarized stars.}
   \end{figure} 
   
It has long been known\cite{Hall49, Hiltner49, Davis51} that weakly aligned, nonspherical dust grains in the interstellar medium preferentially scatter starlight with electric field orientations parallel to the long axis of the grains. The magnitude of this effect varies linearly with heliocentric distance for a given sightline\cite{Fosalba02}. Its optical to near-infrared wavelength dependence may be described by the following, ``Serkowski" empirical model\cite{Serkowski75, Wilking80, Whittet92}:

	\begin{eqnarray}
\frac{P(\lambda)}{P_{\rm max}} = \exp{\left[-(c_1 \lambda_{\rm max} + c_2) \ln^2\left(\frac{\lambda_{\rm max}}{\lambda}\right) \right]}.
	\label{eq:4}
	\end{eqnarray}
	
\noindent Here, the wavelength of peak interstellar polarization $\lambda_{\rm max}$ is typically $\sim 550$ nm, and it is theorized to be related to the mean ISM dust grain size along the line of sight. The constants $c_1$ and $c_2$ describe the width of the polarization spectrum, and they vary from one sightline to another\cite{Wilking82}. Figure \ref{fig:serk} shows interstellar polarization and Serkowski fits for the nearly unpolarized star GJ 9069A and the strongly polarized stars HD 77581 and HD 78344. In addition to the $V$ band data for HD 77581, there exist published $UBGRI$ data\cite{Dolan88} that sample near the peak wavelength of interstellar polarization for this system. Even though HD 77581 is a well-known high mass X-ray binary (= Vela X-1), its polarimetric amplitude of orbital modulation is only roughly $\pm 0.1\%$, which is near the limit of accuracy with GPI.

The near-infrared polarizations measured by GPI for the strongly polarized stars are consistent with interstellar polarization. Therefore, it appears that the absolute accuracy in linear polarization measurements obtainable with GPI is $\sim 0.1\%$ or less (Figure \ref{fig:serk}), which is far below the expected polarization for spatially resolved, circumstellar material. GPI's ability to accurately and precisely measure the polarization of circumstellar dust-scattered light will be limited by contrast and our ability to measure the total intensity $I$, not by any systematic biases in the measurement of $Q$ and/or $U$. 

Interestingly, this level of accuracy is similar to the expected polarization of thermal emission from rapidly rotating, oblate exoplanets or those with non-uniform cloud cover\cite{Sengupta10, Marley11}. Thus, GPI may not only directly image scattered light from tenuous circumstellar material, but it may also probe asymmetry in thermal emission from young exoplanets.
\section{Suppression of Unpolarized Light}
\label{sec:sup}

\begin{figure}
   \begin{center}
   \begin{tabular}{c}
   \includegraphics[width=15 cm]{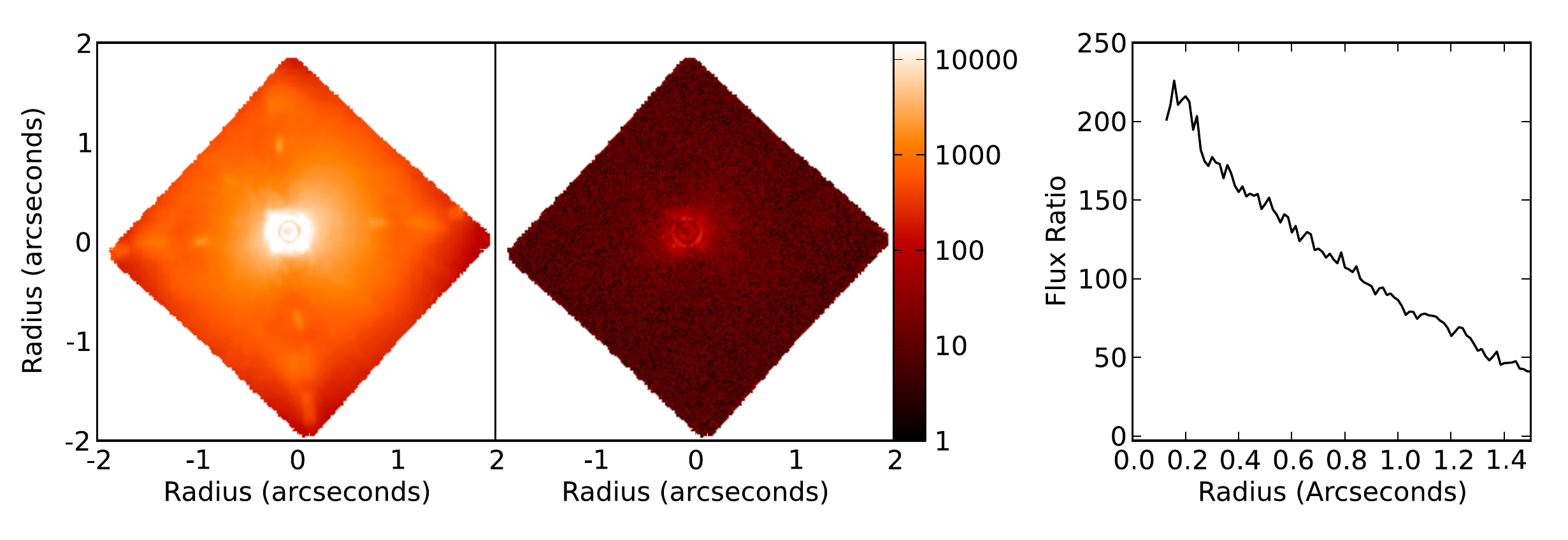}
   \end{tabular}
   \end{center}
   \caption[example] 
   { \label{fig:pol_imgs} 
   \textbf{Left:} The total intensity (left) and linear polarized intensity (right) of the star HD 118666. Residual polarized intensity can be due to both instrumental polarization and noise (e.g. photon noise, read out noise, etc.) in the image, because the linear polarized intensity is a positive defined quantity. \textbf{Right:} The ratio between the mean total intensity and mean linear polarized intensity as a function of radius. At small radii the suppression of unpolarized light appears to be dominated by instrumental polarization. At larger radii the suppression appears to be dominated by photon noise. 
}
   \end{figure}

With the ultimate goal of detecting faint circumstellar material in mind, we observed the unpolarized standard HD 118666 ($P=0.07 \pm 0.035\%$ in $V$ band\cite{Mathewson70}) during a March 2014 commissioning run to assess GPI's ability to cancel out unpolarized light.  HD 118666 was observed in GPI's closed-loop coronagraph mode in $J$, $H$ and $K1$ filters; however, here we present the $H$ band data only, leaving the analysis of the other two bands for future work (M. Perrin et al., 2014, in preparation). The $H$ band observations consisted of two contiguous sets of four HWP orientations, each with 60 s exposures, for a total of eight minutes of integration. Total intensity (Stokes $I$) and linear polarized intensity images were generated using standard reduction techniques\cite{Perrin10} (Figure~\ref{fig:pol_imgs}, left panel). The ratio of the mean total intensity and mean linear polarized intensity at each radius provides an estimate of the level to which unpolarized speckles have been suppressed (Figure~\ref{fig:pol_imgs}, right panel). We find that within $\sim 0.3"$ of the central obscuration, unpolarized light is suppressed by over a factor of 200. The maximum cancellation reached is a factor of $\sim~226$, which is in good agreement with the measured instrumental linear polarization of $I \rightarrow P_{\rm IP} = 0.4353 \pm 0.0075\% $. Thus, it appears that our ability to suppress unpolarized light may be limited by the instrumental polarization, which was not yet taken into account in the data processing for these cubes. It is work in progress to ascertain whether improved starlight suppression can be achieved once the instrumental polarization calculation derived above is incorporated into the GPI Data Reduction Pipeline. The  cancellation ratio decreases with increasing separation from the star; factors such as photon noise and detector read noise become more relevant as the overall apparent surface brightness decreases. Note that at separations outside of $\sim$0.5'', noise models show that the observed cancellation ratio (equivalently, the observed floor in polarized intensity $P$) is consistent with GPI reaching the photon noise floor set by the stellar PSF halo.  

\section{CONCLUSION}
We have presented commissioning observations of the Gemini Planet Imager's polarimetry mode with the aim of instrument calibration. Stellar polarization may be distinguished from instrumental polarization due to the parallactic angle dependence of stellar polarization in the reference frame of the instrument. Long duration observations of the nearly unpolarized star $\beta$ Pic, through a parallactic angle range of $\sim 92^\circ$, are consistent with the following instrumental polarization: $I \rightarrow (Q_{\rm IP}, U_{\rm IP}, P_{\rm IP}, V_{\rm IP}) = (-0.037 \pm 0.010\%, +0.4338 \pm 0.0075\%, 0.4354 \pm 0.0075\%, -6.64 \pm 0.56\%)$. These values are broadly consistent with optical modeling of the instrument. We determine an upper limit to linear polarization introduced by the coronagraph occulter of $\Delta (Q, U) < 0.17\%$ with $95\%$ confidence. A similar level of absolute accuracy in polarimetric measurements with GPI is obtained by fitting multiwavelength observations of strongly polarized stars to a well-established, empirical model of interstellar polarization.

Together with GPI's advanced adaptive optics and coronagraphic starlight suppression, the $\sim 0.1\%$ accuracy in GPI linear polarimetric measurements will enable groundbreaking, spatially resolved investigations of tenuous circumstellar material. GPI's imaging polarimetry mode was designed and its calibration program planned with a goal that statistical measurement errors and not systematic biases should dominate the noise budget\cite{Perrin10}. This led to a requirement that instrumental linear polarization be calibrated to better than 0.3\%. The observations presented here, though preliminary, demonstrate on-sky calibration performance meeting and exceeding that requirement. Further work will extend this into a comprehensive, multi-wavelength characterization of GPI's performance as an imaging polarimeter.  

Polarimetry has been a popular mode of GPI so far, with extensive use in the shared-risk early science period for studies of debris disks, transitional disks, protoplanetary disks, post-main-sequence stellar mass loss, and more. Observations of circumstellar disks will surely continue to be one of the main thrusts of GPI's overall scientific program.  Precise calibrations also open up the possibility of observing the weak polarization signatures from young exoplanets with spatially asymmetric thermal emission, either by oblateness engendered by rapid rotation or by patchy clouds\cite{Marley11}.

\acknowledgments     
 
SJW performed this work in part under contract with the California Institute of Technology (Caltech) funded by NASA through the Sagan Fellowship Program. This work was supported in part by the University of California Lab Research Program 09-LR-01-118057-GRAJ and NSF AST-0909188. Work at LLNL was performed under the auspices of DOE under contract DE-AC52-07NA27344. This research was supported in part by the National Science Foundation Science and Technology Center for Adaptive Optics, managed by the University of California at Santa Cruz under cooperative agreement No. AST - 9876783. We also acknowledge support from the Natural Science and Engineering Council of Canada. The Gemini Observatory is operated by the Association of Universities for Research in Astronomy, Inc., under a cooperative agreement with the NSF on behalf of the Gemini partnership: the National Science Foundation (United States), the National Research Council (Canada), CONICYT (Chile), the Australian Research Council (Australia), Minist\'{e}rio da Ci\^{e}ncia, Tecnologia e Inova\c{c}\~{a}o (Brazil) and Ministerio de Ciencia, Tecnolog\'{i}a e Innovaci\'{o}n Productiva (Argentina).


\bibliography{report}   
\bibliographystyle{spiebib}   

\end{document}